\documentclass[twocolumn]{aastex63}

\usepackage{amsmath}
\usepackage{color}

\received{\today}
\revised{\today}
\accepted{\today}
\submitjournal{ApJ}

\shorttitle{The cosmic \hi\ gas mass density in galaxies at $z\approx 7$}
\shortauthors{Heintz et al.}

\newcommand{\hi}{H\,{\sc i}}
\newcommand{\cii}{C\,{\sc ii}}
\newcommand{\ci}{C\,{\sc i}}

\begin{document}

\title{The ALMA REBELS Survey: The Cosmic \hi\ Gas Mass Density in Galaxies at $z\approx 7$}

\correspondingauthor{K.~E.~Heintz}
\email{keheintz@nbi.ku.dk}

\author[0000-0002-9389-7413]{K.~E.~Heintz}
\affiliation{Cosmic Dawn Center (DAWN), Denmark}
\affiliation{Niels Bohr Institute, University of Copenhagen, Jagtvej 128, DK-2200 Copenhagen N, Denmark}

\author[0000-0001-5851-6649]{P.~A.~Oesch}
\affiliation{University of Geneva, Department of Astronomy, Chemin Pegasi 51, 1290 Versoix, Switzerland}
\affiliation{Cosmic Dawn Center (DAWN), Denmark}
\affiliation{Niels Bohr Institute, University of Copenhagen, Jagtvej 128, DK-2200 Copenhagen N, Denmark}

\author[0000-0002-6290-3198]{M.~Aravena}
\affiliation{Nucleo de Astronomia, Facultad de Ingenieria y Ciencias, Universidad Diego Portales, Av. Ejercito 441, Santiago, Chile}

\author[0000-0002-4989-2471]{R.~J.~Bouwens}
\affiliation{Leiden Observatory, Leiden University, NL-2300 RA Leiden, Netherlands}

\author[0000-0001-8460-1564]{P.~Dayal}
\affiliation{Kapteyn Astronomical Institute, University of Groningen, P.O. Box 800, 9700 AV Groningen, The Netherlands}

\author[0000-0002-9400-7312]{A.~Ferrara}
\affiliation{Scuola Normale Superiore, Piazza dei Cavalieri 7, I-56126 Pisa, Italy}

\author[0000-0001-7440-8832]{Y.~Fudamoto}
\affiliation{Waseda Research Institute for Science and Engineering, Faculty of Science and Engineering, Waseda University, 3-4-1 Okubo, Shinjuku, Tokyo 169-8555, Japan}
\affiliation{National Astronomical Observatory of Japan, 2-21-1, Osawa, Mitaka, Tokyo, Japan}

\author[0000-0002-9231-1505]{L.~Graziani}
\affiliation{Dipartimento di Fisica, Sapienza, Università di Roma, Piazzale Aldo Moro 5, I-00185 Roma, Italy}
\affiliation{INAF/Osservatorio Astronomico di Roma, via Frascati 33, I-00078 Monte Porzio Catone, Roma, Italy}

\author[0000-0003-4268-0393]{H.~Inami}
\affiliation{Hiroshima Astrophysical Science Center, Hiroshima University, 1-3-1 Kagamiyama, Higashi-Hiroshima, Hiroshima 739-8526, Japan}

\author[0000-0002-2906-2200]{L.~Sommovigo}
\affiliation{Scuola Normale Superiore, Piazza dei Cavalieri 7, I-56126 Pisa, Italy}

\author[0000-0001-8034-7802]{R.~Smit}
\affiliation{Astrophysics Research Institute, Liverpool John Moores University, 146 Brownlow Hill, Liverpool L3 5RF, United Kingdom}

\author[0000-0001-7768-5309]{M.~Stefanon}
\affiliation{Departament d'Astronomia i Astrof\'isica, Universitat de Val\`encia, C. Dr. Moliner 50, E-46100 Burjassot, Val\`encia, Spain}

\author[0000-0001-8426-1141]{M.~Topping}
\affiliation{Steward Observatory, University of Arizona, 933 N Cherry Ave, Tucson, AZ 85721, USA}

\author[0000-0002-7129-5761]{A.~Pallottini}
\affiliation{Scuola Normale Superiore, Piazza dei Cavalieri 7, I-56126 Pisa, Italy}

\author[0000-0001-5434-5942]{P.~van~der~Werf}
\affiliation{Leiden Observatory, Leiden University, NL-2300 RA Leiden, Netherlands}

\begin{abstract}
The neutral atomic gas content of individual galaxies at large cosmological distances has until recently been difficult to measure due to the weakness of the hyperfine \hi\ 21-cm transition. Here we estimate the \hi\ gas mass of a sample of main-sequence star-forming galaxies at $z\sim 6.5 - 7.8$ surveyed for [\cii]$-158\mu$m emission as part of the Reionization Era Bright Emission Line Survey (REBELS), using a recent calibration of the [\cii]-to-\hi\ conversion factor. We find that the \hi\ gas mass excess in galaxies increases as a function of redshift, with an average of $M_{\rm HI} / M_\star \approx 10$, corresponding to \hi\ gas mass fractions of $f_{\rm HI} = M_{\rm HI} / (M_\star + M_{\rm HI}) = 90\%$, at $z\approx 7$.
Based on the [\cii]-$158\mu$m luminosity function (LF) derived from the same sample of galaxies, we further place constraints on the cosmic \hi\ gas mass density in galaxies ($\rho_{\rm HI}$) at this redshift, which we measure to be $\rho_{\rm HI} =  7.1^{+6.4}_{-3.0} \times 10^{6}\,M_{\odot}\,{\rm Mpc^{-3}}$. This estimate is substantially lower by a factor of $\approx 10$ than that inferred from an extrapolation of damped Lyman-$\alpha$ absorber (DLA) measurements, and largely depend on the exact [\cii] LF adopted. However, we find this decrease in $\rho_{\rm HI}$ to be consistent with recent simulations and argue that this apparent discrepancy is likely a consequence of the DLA sightlines predominantly probing the substantial fraction of \hi\ gas in high-$z$ galactic halos, whereas [\cii] traces the \hi\ in the ISM associated with star formation. 
We make predictions for this build-up of neutral gas in galaxies as a function of redshift, showing that at $z\gtrsim 5$ only $\approx 10\%$ of the cosmic \hi\ gas content is confined in galaxies and associated with the star-forming ISM. 
\end{abstract}


\section{Introduction} \label{sec:intro}

In the modern cosmological framework, the first epoch of galaxy formation and reionization is initiated by the infall of neutral, pristine gas \citep{Dayal18}. Through the hierarchical formation of structure, the accretion and gravitational collapse of a significant fraction of this gas allows it to condense into gaseous "proto-galactic" haloes. This gas can internally cool and form dense molecular clouds that will eventually ignite and start producing stars. The neutral atomic hydrogen (\hi) gas content is therefore a fundamental component in the formation and evolution of the first generation of galaxies.

In the local universe, the \hi\ gas component in galaxies can be directly measured via the hyperfine \hi\ 21-cm line emission \citep{Zwaan05,Walter08,Leroy08}. However, due to the weakness of the line, this feature can only be detected for individual galaxies out to modest cosmological distances \citep[the most distant being located at $z=0.367$;][]{Fernandez16}. Recent advances measuring the integrated 21-cm signal from thousands of galaxies have pushed this technique out to $z\approx 1$, providing a census of the average \hi\ gas content of star-forming galaxies at this redshift \citep{Chowdhury20,Chowdhury21}. 

Alternatively, \hi\ can be detected in absorption via the 21-cm or damped Lyman-$\alpha$ absorption (DLA) features imprinted from foreground galaxies towards bright background quasars \citep{Peroux03}. This has enabled independent measurements of the cosmic \hi\ gas mass density from $z\sim 0-5$ \citep{Prochaska09,Noterdaeme12,Neeleman16}. However, there are some uncertainties in how these DLA systems trace the star-forming galaxy population, with recent results showing that they predominantly probe the extended \hi\ in the halos of galaxies \citep{Neeleman19}. These extended \hi\ gas reservoirs therefore does not represent the neutral gas associated with star formation.
Further, this approach of measuring \hi\ is naturally limited to $z\lesssim 5$ due to the onset of the Lyman-$\alpha$ forest and the increasing Gunn-Peterson effect, significantly hindering measurements of DLA features in intervening galaxies at these early epochs \citep[but see e.g.][]{Banados19}.

Due to these limitations, the most promising approach to obtain a census of the overall \hi\ gas content of galaxies beyond $z\gtrsim 5$ may be to use a suitable tracer of this cold neutral gas-phase. Similar approaches based on the emission from carbon monoxide \citep[CO;][]{Tacconi10,Tacconi18}, neutral atomic carbon \citep[\ci;][]{Papadopoulos04,Walter11,Valentino18}, or dust \citep{Magdis12,Scoville17} is commonly used to infer the molecular gas mass of galaxies through their connection to molecular hydrogen, H$_2$ \citep{Bolatto13}. Similarly, it is of great interest to establish a feasible, alternative proxy for \hi. 

The bright infrared fine-structure transition of ionized carbon [\cii]-158\,$\mu$m is a promising tracer of gas in the most distant galaxies \citep{Carilli13}, even reaching into the epoch of reionization at $z\gtrsim 6$ \citep[e.g.,][]{Smit18,Matthee19,Hashimoto19,Schouws22}. 
The relatively low ionization potential of C (${\rm IP} = 11.26$\,eV) causes a significant fraction of carbon in the neutral gas-phase to be ionized to C$^+$ due to the permeating interstellar radiation field. [\cii] emission has indeed been observed to originate partly from the neutral gas-phase in the Milky Way and nearby galaxies 
\citep[e.g.,][]{Madden93,Madden97,Pineda14,Croxall17}, with growing evidence in high-redshift galaxies as well \citep[e.g.,][]{Novak19,Meyer22}.
However, some contributions to the total emission are also expected from the ionized gas and molecular regions \citep[e.g.,][]{Olsen15,Vallini17,Pallottini19}, though simulations of galaxies at $z\approx 6$ predict [\cii] to predominantly originate from the neutral ISM \citep{Katz17,RamosPadilla22}. In a recent study, \citet{Heintz21} determined an empirical relation, specifically linking the [\cii]-158\,$\mu$m line emission to the total \hi\ gas mass in galaxies. This [\cii]-to-\hi\ conversion factor (here denoted $\beta_{\rm [CII]}$) appears universal in galaxies from $z\sim 6.5$ to $z\sim 0$, based on simulations and direct observations. 

Here we apply this $\beta_{\rm [CII]}$ conversion factor to infer the \hi\ gas content of a statistical sample of UV-selected star-forming galaxies at $z\sim 6.5 - 7.8$, surveyed for [\cii]$-158\mu$m emission as part of the Reionization Era Bright Emission Line Survey \citep[REBELS;][]{Bouwens22}. The molecular gas content of these galaxies will be studied in a separate work (Aravena et al., in prep.).
The stellar masses and star-formation rates (SFRs) of these galaxies are representative of the high-mass end of the main-sequence at this epoch \citep[e.g.,][]{Topping22}. 
In Sect.~\ref{sec:met} we introduce the sample and describe the methodology. In Sect.~\ref{sec:res} we present our results. In Sect.~\ref{sec:rhohi} we discuss the implications of the inferred global \hi\ gas content for the evolution of galaxies in the reionization epoch and we conclude on our work in Sect.~\ref{sec:conc}. 

Throughout the paper we assume the concordance $\Lambda$CDM cosmological model with $\Omega_{\rm m} = 0.315$, $\Omega_{\Lambda} = 0.685$, and $H_0 = 67.4$\,km\,s$^{-1}$\,Mpc$^{-1}$ \citep{Planck18}, and adopt a \citet{Chabrier03} initial mass function (IMF). 


\section{Sample and methodology} \label{sec:met}

REBELS targeted 40 galaxies at $z\sim 6.5 - 9.0$ selected from fields with deep photometric auxillary data, as presented in \citet{Bouwens22}. 
We adopt the total SFRs (UV + IR) and stellar masses presented by \citet{Topping22} here. They are in the range 
$M_\star = 6.5\times 10^{8} - 3.1\times 10^{10}\,M_\odot$ and ${\rm SFRs} = 10 - 200\,M_\odot$\,yr$^{-1}$,
determined by modelling the spectral energy distribution (SED) for each source with {\tt Prospector} \citep{Leja17} assuming a non-parametric star-formation history. The inferred stellar masses are generally higher than for models assuming a constant star-formation history.
Of the surveyed galaxies, 23 were detected in [\cii], with luminosities ranging from $L_{\rm [CII]} = 1.3\times 10^{8}\,L_\odot$ to $1.7\times 10^{9}\,L_\odot$ (Schouws et al. in prep.). 
Other recent results from the REBELS survey include constraints on the dust properties, masses and temperatures \citep{Sommovigo22,Ferrera22,Dayal22,Iname22}, sizes and morphologies of the [\cii] emission components \citep{Fudamoto22} and the efficiency of Ly$\alpha$ transmission \citep{Endsley22} of galaxies during the epoch of reionization.

To apply the metallicity-dependent [\cii]-to-\hi\ conversion factor derived by \citet{Heintz21} we first need to infer the gas-phase metallicity for each galaxy in the sample. We adopt the fundamental $M_\star$-SFR-metallicity relation from \citet{Curti20}, parametrized as
\begin{equation}
    Z(M_\star,{\rm SFR}) = Z_0 - \gamma / \beta \log (1 + (M_\star/M_0({\rm SFR}))^{-\beta})
\end{equation}
where $M_0({\rm SFR}) = 10^{m_0}\times {\rm SFR}^{m_1}$. We adopt their best-fitting parameters considering the total SFR: $Z_0 = 8.779$, $m_0 = 10.11$, $m_1 = 0.56$, $\gamma = 0.31$, and $\beta = 2.1$. This yields a range in metallicities for the REBELS sample of $12+\log({\rm O/H}) = 8.0-8.6$ ($\log Z/Z_\odot = -0.7$ to $-0.1$), with a mean and median value of 8.40 and 8.39, respectively. We caution, however, that this fundamental plane relation is not well-established at $z\gtrsim 3$. These estimates should therefore only be taken as indicative of the actual metallicities. The recent calibrations provided by the {\tt Astraeus} simulation framework \citep{Hutter21,Ucci21} indeed seem to indicate overall lower metallicities at the same SFRs and stellar masses, but to be conservative we adopt the empirical relation from \citet{Curti20}, effectively providing a lower bound on $M_{\rm HI}$. Our metallicity estimates are also in good agreement with measurements of individual star-forming galaxies at similar redshifts \citep{Jones20fmr} and follows the expected decrease in metallicity as a function of redshift \citep{Sanders20}. 

Based on the inferred gas-phase metallicity and the measured $L_{\rm [CII]}$ for each galaxy, we estimate the \hi\ gas mass using the $\beta_{\rm [CII]}$ conversion factor 
\begin{equation}
\begin{split}
    \log \beta_{\rm [CII]} = \log M_{\rm HI} / L_{\rm [CII]} = (-0.87\pm 0.09) \times \\ \log(Z/Z_{\odot}) + (1.48\pm 0.12) 
    \label{eq:ciitohi}
\end{split}
\end{equation}
calibrated by \citet{Heintz21}, where $Z/Z_{\odot}$ is the relative solar abundance with $12+\log({\rm O/H})_{\odot} = 8.69$ for $\log (Z/Z_{\odot}) = 0$ \citep{Asplund09}, and $M_{\rm HI}$ and $L_{\rm [CII]}$ are in units of $M_{\odot}$ and $L_{\odot}$, respectively. This calibration is measured directly in galaxies at $z\sim 2-6$ through pencil-beam sightlines from $\gamma$-ray burst (GRB) afterglows and absorption-line spectroscopy. This approach does not assume or require that [\cii] and \hi\ have to be physically associated, though this seems to be evident from the absorption-line spectra. The derived $\beta_{\rm [CII]}$ abundance ratio simply provides a measure of the total ``column'' [\cii] luminosity, irrespective of the phase it originates from, and relates it to the \hi\ column density in the same sightline. This ratio is, however, assumed to be representative of the galaxy average. A similar methodology was used to derive an accurate [\ci]-to-H$_2$ conversion factor in equivalent high-$z$ absorption-selected galaxies \citep{HeintzWatson20}. The $\beta_{\rm [CII]}$ calibration was further supported and observed to show remarkably consistency with local dwarf galaxies at $z\sim 0$ from the {\it Herschel} Dwarf Galaxy Survey \citep{Madden13}, for which [\cii] luminosities and \hi\ gas masses have been measured from direct 21-cm observations \citep{RemyRuyer14,Cormier15}, and simulations of galaxies at similar redshifts. The metallicity-dependent [\cii]-to-\hi\ $\beta_{\rm [CII]}$ conversion factor thus appears to be universal across redshifts. With a systematic uncertainty of 0.3 dex on the inferred metallicities, an additional 0.25 dex uncertainty will be propagated to $\log M_{\rm HI}$ following this relation, which we assume throughout. We also note that if the actual metallicities are potentially underestimated as described above, the estimates of $M_{\rm HI}$ should effectively be treated as lower limits. The observed anti-correlation with metallicity of the $\beta_{\rm [CII]}$ conversion factor is primarily a consequence of the increasing C/H abundance with increasing gas-phase metallicity. 

\begin{figure}[t!]
\centering
\includegraphics[width=9.5cm]{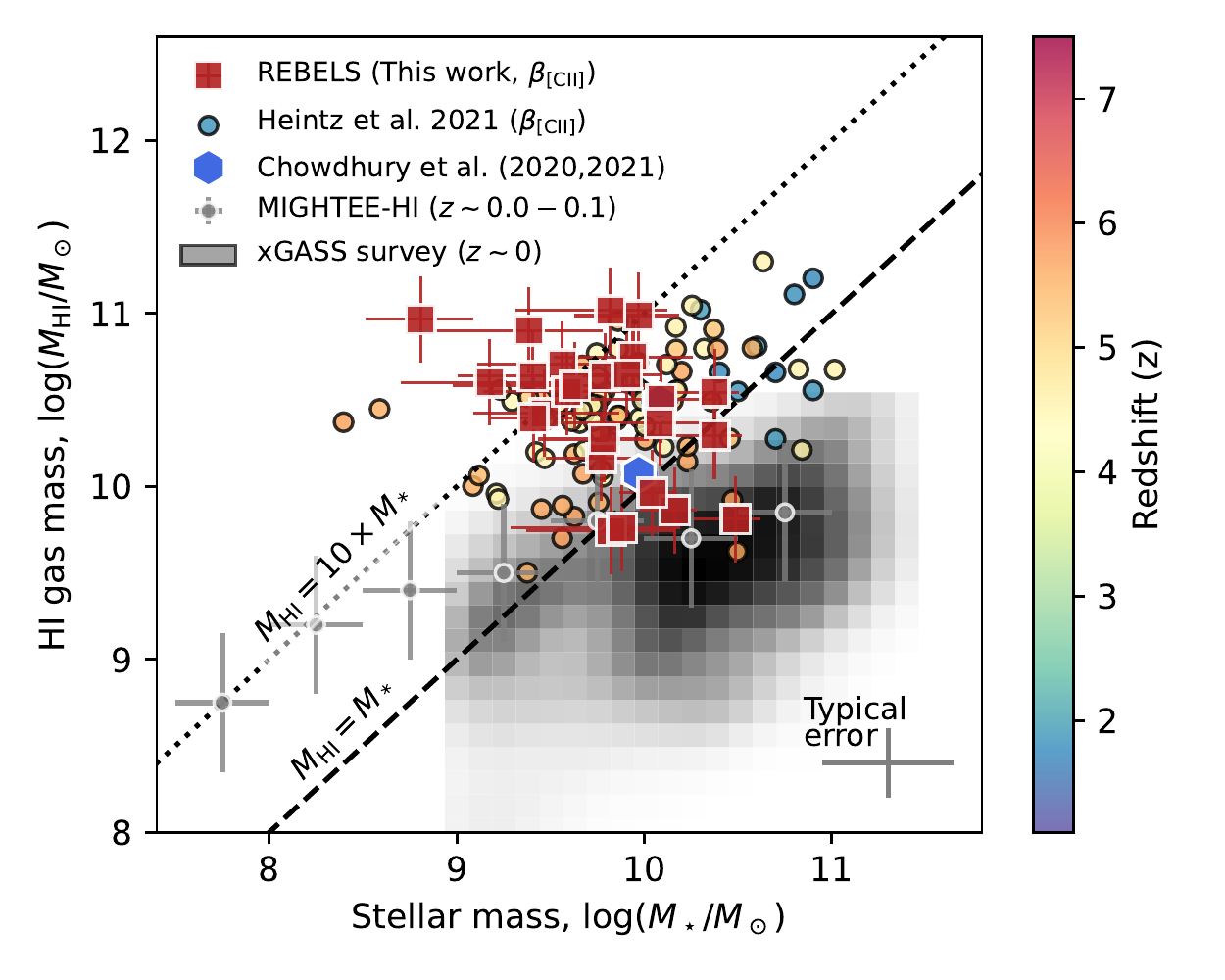}
\caption{\hi\ gas and stellar mass distributions of the REBELS sample galaxies at $z\sim 6.5-7.8$ (red squares). The inferred \hi\ gas and stellar masses based on the $\beta_{\rm [CII]}$ calibration of other high-redshift galaxy samples, color-coded as a function of redshift, are shown for comparison \citep[see text and][for references]{Heintz21}, in addition to average \hi\ gas mass excess inferred by \citet{Chowdhury20,Chowdhury21} from a stack of the 21-cm signal from galaxies at $z\approx 1-1.3$ (blue hexagons). The uncertainties on the REBELS sources are marked individually, and typical uncertainties for the other samples relying on $\beta_{\rm [CII]}$ are shown as well. The xGASS sample represents galaxies at $z\sim 0$ with direct 21-cm \hi\ observations. Diagonal lines mark $M_{\rm HI}/M_\star = 1 - 10$. }
\label{fig:mhimstar}
\end{figure}

\section{Results} \label{sec:res}

We derive \hi\ gas masses in the range $M_{\rm HI} = 5.5\times 10^{9}\,M_\odot$ to $1.0\times 10^{11}\,M_\odot$ for the REBELS galaxy sample at $z\sim 6.5 - 7.8$. We find $M_{\rm HI} > M_\star$ for the majority of systems, with mean and median values of $M_{\rm HI} / M_\star = 13.3$ and 6.4, respectively, corresponding to \hi\ gas mass fractions of $f_{\rm HI} = M_{\rm HI} / (M_\star + M_{\rm HI}) = 86\%$ and $93\%$. These estimates are in excellent agreement with the gas fractions of galaxies at $z\gtrsim 6$ predicted by the {\tt Astraeus} simulations \citep{Ucci21}. The $M_{\rm HI}$ vs. $M_\star$ distributions are presented in Fig.~\ref{fig:mhimstar}. For comparison, we overplot as 2D density contours in the $M_{\rm HI}-M_\star$ parameter space the extended GALEX Arecibo SDSS Survey (xGASS) catalog of galaxies at $z\sim 0$ with direct \hi\ 21-cm observations \citep{Catinella18}. We also include the median measurements of the Early Science galaxies from the MIGHTEE \hi\ emission project covering $z=0.0-0.084$ \citep{Maddox21}, and the average $M_{\rm HI}$ and $M_\star$ measurements by \citet{Chowdhury20,Chowdhury21}, from stacking of the 21-cm signal of galaxies at $z\sim 1-1.3$. 
We further show the measurements from \citet{Heintz21} based on the $\beta_{\rm [CII]}$ conversion factor of galaxies in the range $z\sim 2 - 6$ surveyed for [\cii] emission from the samples by \citet{Zanella18}, \citet{Capak15} and the ALPINE-[\cii] survey \citep{LeFevre20,Bethermin20,Faisst20}. The majority of galaxies at $z\gtrsim 2$ show $M_{\rm HI} > M_\star$ at all stellar masses. Galaxies at $z\lesssim 1$ mostly contain $M_{\rm HI} \lesssim M_\star$ at large stellar masses ($M_\star \gtrsim 10^{9}\,M_{\odot}$), but show increasing $M_{\rm HI} / M_\star$ ratios at lower stellar masses. 

\begin{figure}[t!]
\centering
\includegraphics[width=8.9cm]{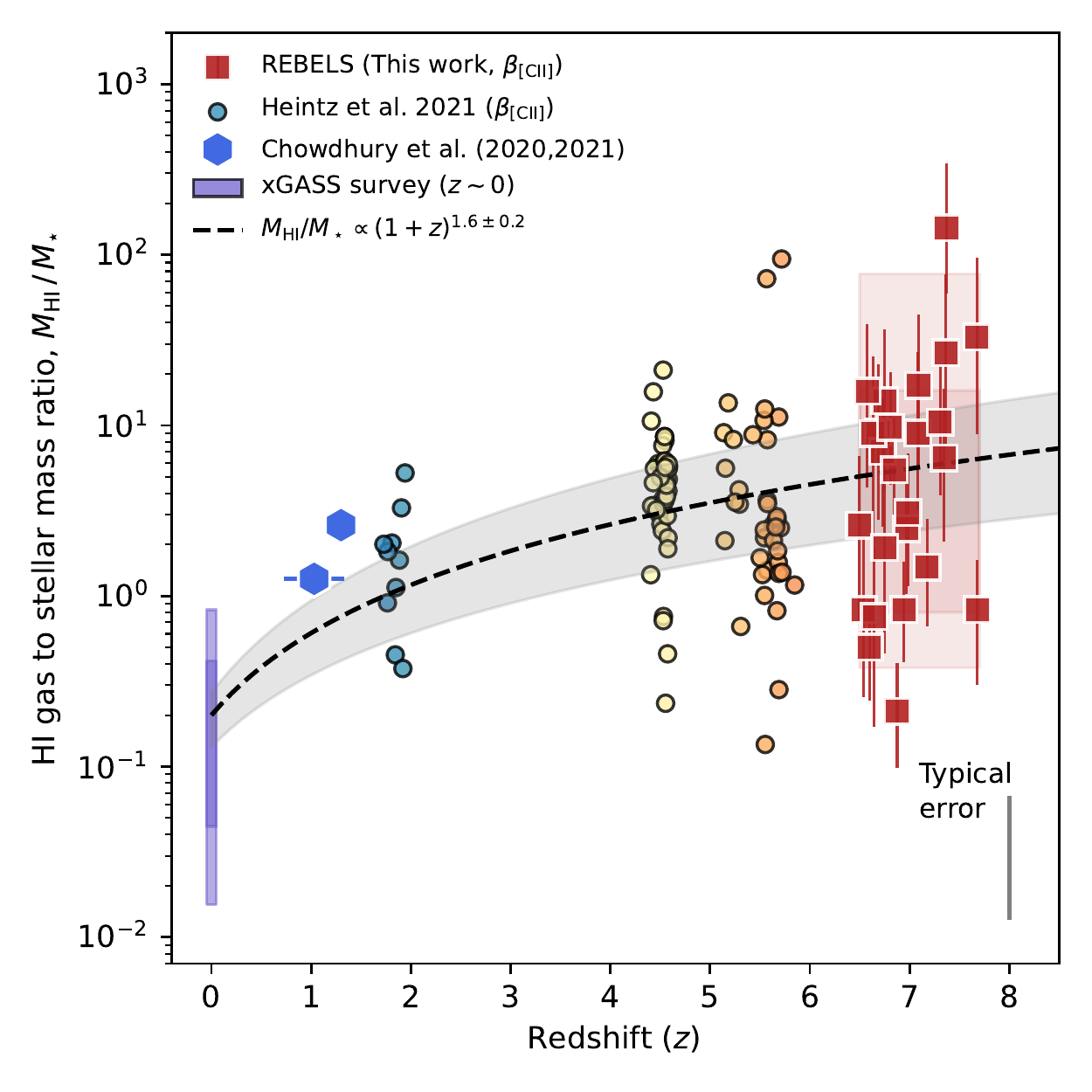}
\caption{Redshift evolution of $M_{\rm HI}/M_\star$. The symbol notation follows Fig.~\ref{fig:mhimstar}. The uncertainties on the REBELS sources (red squares) are marked individually, and typical uncertainties for the other samples relying on $\beta_{\rm [CII]}$ are shown as well. The red-and purple-colored boxes mark vertically the 68\% and 95\% distributions of the REBELS and xGASS samples, respectively, and horizontally spans the full redshift distribution. The average $M_{\rm HI}/M_\star$ ratio is observed to gradually increase from $z\sim 0$ to $z\sim 7$, following approximately $M_{\rm HI}/M_{\star} \propto (1+z)^{1.6\pm 0.2}$ (black dashed line).}
\label{fig:mhimstarz}
\end{figure}

The elevated \hi\ gas mass fractions $M_{\rm HI} / M_\star$ observed in the high-redshift galaxy samples are consistent with a gradual increase as a function of redshift, as illustrated in Fig.~\ref{fig:mhimstarz}. In this figure we compare our measurements to the same local and high-redshift galaxy samples as compiled for Fig.~\ref{fig:mhimstar} and described above, relying on a combination of $M_{\rm HI}$ inferred via direct 21-cm observations and the $\beta_{\rm [CII]}$ conversion. We find that the \hi\ gas mass excess increases as a function of redshift, following approximately $M_{\rm HI} / M_{\star} = 0.2\times (1+z)^{1.6}$, anchored to the median value inferred for xGASS galaxy sample at $z=0$. This relation is found to closely follow the redshift evolution of the specific SFR (sSFR) recently measured by \citep{Topping22}. 

To make a crude prediction for the average fraction of \hi\ in the ISM out of the total baryonic matter content by mass (i.e. $M_{\rm bar,tot} = M_{\rm HI} + M_{\rm H_2} + M_\star$) in these galaxies, we adopt an average [\cii]-to-H$_2$ ratio of $\alpha_{\rm [CII]} = 18$ \citep{Vizgan22}. This yields a dominant \hi\ component, with $M_{\rm HI}/M_{\rm bar,tot} = 60\%$. If we instead assume $\alpha_{\rm [CII]} = 30$ \citep{Zanella18}, we find a slightly lower, but still dominant \hi\ fraction of $M_{\rm HI}/M_{\rm bar,tot} = 55\%$. This suggests that \hi\ dominates the baryonic matter content of galaxies in the epoch of reionization at $z\approx 7$. We further note that the predicted total gas masses $M_{\rm gas} = M_{\rm HI} + M_{\rm H_2}$ of these high-$z$ galaxies follow a relation on global scales similar to the local Kennicutt-Schmidt law \citep[][]{Kennicutt98} of surface densities $({\rm SFR}/M_\odot\,{\rm yr}^{-1}) \propto (M_{\rm gas}/M_\odot)^{n}$ with $n=1.2-1.4$, potentially hinting at the universality of this law even at $z\approx 7$.

\begin{figure*}[!t]
\centering
\includegraphics[width=0.85\textwidth]{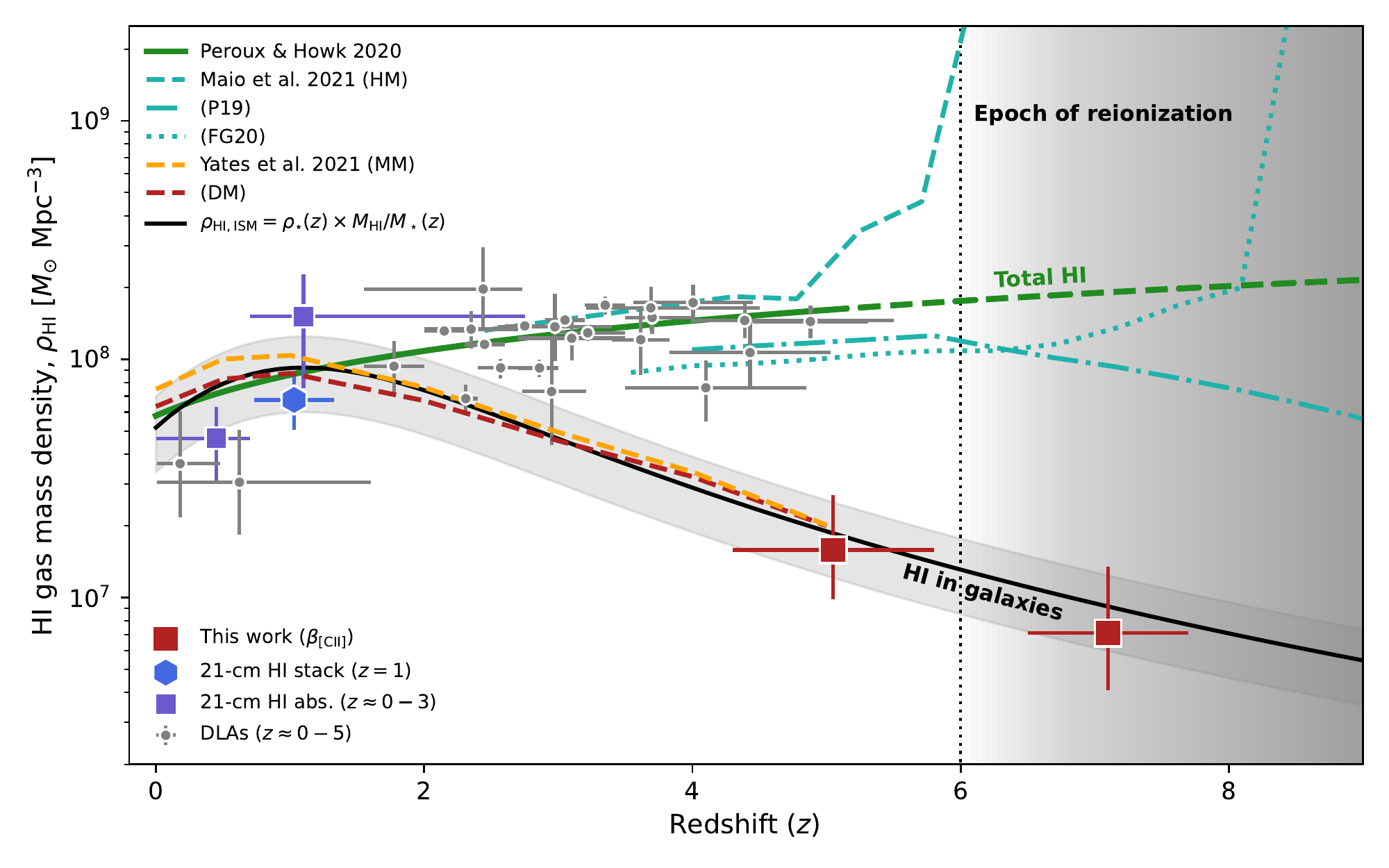}
\caption{Redshift evolution of the cosmic \hi\ gas mass density in galaxies, $\rho_{\rm HI}$. Measurements for this work (red squares) are based on the estimated [\cii] luminosity densities at $z\sim 5$ and $z\sim 7$, converted to $\rho_{\rm HI}$. These results are largely dependent on the exact [\cii] LF and may include significant systematic biases (see text). The blue hexagon shows the average measurement of galaxies at $z\sim 1$ from stacking of the 21-cm \hi\ signal \citep{Chowdhury20} and purple squares show the results from a survey of intervening 21-cm \hi\ absorption systems covering $0 < z < 0.69$ and $0.69 < z < 2.74$ \citep{Grasha20}. The DLA measurements compiled by \citet{Walter20} are shown by the gray dots, corrected by 20\% to account for the contribution to \hi\ from sub-DLAs and Lyman-limit systems. The green curve shows the best-fit evolution of $\rho_{\rm HI,DLA}(z)$ by \citet{Peroux20} based on DLA measurements and extrapolated beyond $z\gtrsim 5$ (shown as the dashed curve). The cyan curves show $\rho_{\rm HI,tot}(z)$ predicted by hydrodynamical simulations using various prescriptions \citep[HM, P19, and FG20, see][and main text]{Maio22}. The orange and red dashed curves show the predictions for the evolution of \hi\ inside galaxies from the default and modified model (DM, MM) of the L-Galaxies simulations by \citet{Yates21}. The black solid line shows the combined redshift evolution of the stellar mass density $\rho_\star(z)$ determined by \citet{Walter20} and the \hi\ gas excess $M_{\rm HI}/M_\star$ derived in this work (see Sect.~\ref{sec:res} and Fig.~\ref{fig:mhimstarz}), with the gray-shaded region marking the $\approx 35\%$ uncertainty on this evolutionary curve. The approximate end of the epoch of reionization is marked at $z\sim 6$.}
\label{fig:rhohi}
\end{figure*}

\section{The cosmic \hi\ gas mass density in galaxies at $z\approx 7$} \label{sec:rhohi}

We now consider the volume-averaged \hi\ gas mass density ($\rho_{\rm HI}$) in galaxies, related to the cosmological \hi\ density $\Omega_{\rm HI}$ via $\rho_{\rm HI} = \Omega_{\rm HI} \times \rho_{\rm c}$, where $\rho_{\rm c}$ is the critical density of the universe\footnote{$\rho_{\rm c}$ is determined to be $\rho_{\rm c} = 1.26\times 10^{11}\,M_\odot$\,Mpc$^{-3}$ in the adopted $\Lambda$CDM cosmological paradigm \citep{Planck18}.}. This quantity has been constrained through various different approaches: at $z\sim 0$, extensive galaxy surveys of the 21-cm \hi\ line emission has been used to infer the \hi\ mass function, providing an estimate of $\rho_{\rm HI}$ \citep[e.g.,][]{Zwaan05,Braun12,Jones18HI}. 

At intermediate redshifts ($z\sim 0.4 - 1$), $\rho_{\rm HI}$ can be constrained by stacking the 21-cm signal from a large ensemble of galaxies \citep[e.g.,][]{Lah07,Kanekar16,Chowdhury20} or via observations of the 21-cm feature in absorption toward radio bright continuum sources \citep[e.g.,][]{Grasha20,Allison21}. At $z\gtrsim 2$ and up to $z\approx 5$, the \hi\ gas mass density has primarily been constrained via the Lyman$-\alpha$ feature from intervening absorbers toward bright background quasars \citep[see][and references therein]{Peroux20}. These measurements, however, only constrain the fraction of \hi\ gas in DLAs, $\rho_{\rm HI, DLA}$, which may not be representative of the \hi\ gas associated with star formation in galaxies at high redshifts, even though these system contain the bulk ($>85\%$) of the \hi\ at all redshifts \citep[][]{Zwaan05,OMeara07,Noterdaeme12}. We will address this point further below. 

Following \citet{Heintz21}, we determine $\rho_{\rm HI}$ based on the [\cii]-$158\mu$m luminosity density, $\mathcal{L}_{\rm [CII]}$, at $z\approx 7$. This approach follows a similar methodology to previous works that infer the cosmic H$_2$ gas mass density $\rho_{\rm H_2}(z)$ based on the CO luminosity density \citep[e.g.,][]{Walter14,Decarli19,Riechers19}. 
We adopt the [\cii]-$158\mu$m luminosity function (LF) parameters derived by Oesch et al. (in prep.) based on the REBELS data to estimate the luminosity density as $\mathcal{L}_{\rm [CII]} = \int^{\infty}_{L_{\rm [CII]}} L_{\rm [CII]} \phi(L_{\rm [CII]}) dL_{\rm [CII]}$. The [\cii]-$158\mu$m LF parameters considered in this work were determined by converting the UV LF to a [\cii] LF. As prior, we use the UV LF from \citet{Bouwens21UVLF} and the empirical $L_{\rm UV} - L_{\rm [CII]}$ relation (including an intrinsic dispersion), derived for galaxies at the same redshift. For more details, see Oesch et al., in prep. We integrate over the full [\cii] LF down to $\log (L_{\rm [CII]}/L_{\odot}) = 7.5$, representing approximately the REBELS detection limit \citep[][Schouws et al. in prep.]{Bouwens22}. This yields a [\cii] luminosity density of $\log (\mathcal{L}_{\rm [CII]}/\,L_\odot\,{\rm Mpc}^{-3}) = 4.85^{+0.25}_{-0.20}$ at $z\approx 7$. To infer $\rho_{\rm HI}$, we assume an average gas-phase metallicity of $12+\log({\rm O/H}) = 8.1$ ($\log (Z/Z_\odot) = -0.6$) \citep[][]{Jones20fmr}, to determine the average [\cii]-to-\hi\ conversion factor $\log \beta_{\rm [CII]} = 2.00 \pm 0.13$ at this epoch. This yields a \hi\ gas mass density of $\rho_{\rm HI} = 7.1^{+6.4}_{-3.0} \times 10^{6}\,M_{\odot}\,{\rm Mpc^{-3}}$ in galaxies at $z\approx 7$, also shown in Fig.~\ref{fig:rhohi}.

To be consistent with this work, we re-derive the value of $\mathcal{L}_{\rm [CII]}$, and consequently $\rho_{\rm HI}$, estimated by \citet{Heintz21} at $z\approx 5$. This is based on the recent work by Oesch et al. (in prep.), who find that adopting a similar $L_{\rm UV} - L_{\rm [CII]}$ relation as prior to convert the $z\sim5$ UV LF to a [\cii]$-158\mu$m LF yields results that are in good agreement with the step-wise [\cii] LF points from \citet{Yan20} measured from the ALPINE target sample. However, this [\cii] LF lies significantly below the estimate obtained from the ``serendipitous" $z\sim5$ sources detected in the ALPINE survey as well as from estimates derived via the CO LF, which predict a vast excess of sources with $\log (L_{\rm [CII]}/L_{\odot}) > 9$ compared to the ALPINE target sample \citep[see][]{Yan20,Loiacono21}. This greatly affects the derivation of $\mathcal{L}_{\rm [CII]}$, and thus also $\rho_{\rm HI}$, which we find to be a factor of $\approx 10$ lower than the previous estimate \citep[at $3\sigma$ significance][]{Heintz21}. To be consistent with the REBELS LF, we chose to adopt the step-wise [\cii] LF points derived for the ALPINE target sample by \citet{Yan20}, including the UV LF as prior to better represent the underlying intrinsic galaxy population. For a more detailed discussion see Oesch et al. (in prep.).
With this updated [\cii]$-158\mu$m LF, we estimate $\log (\mathcal{L}_{\rm [CII]}/\,L_\odot\,{\rm Mpc}^{-3}) = 5.37^{+0.19}_{-0.16}$ at $z\approx 5$, from which we now infer $\rho_{\rm HI} = 1.6^{+1.1}_{-0.6} \times 10^{7}\,M_{\odot}\,{\rm Mpc^{-3}}$ as also shown in Fig.~\ref{fig:rhohi}. Overall, we find that $\rho_{\rm HI}$ increases by a factor of $\approx 10$ from $z\approx 7$ to $z=2$, to then turnover and decrease at later cosmic times, reaching $\rho_{\rm HI} = 5\times 10^{7}\,M_{\odot}\,{\rm Mpc^{-3}}$ at $z=0$ \citep{Peroux20,Walter20}.

In Fig.~\ref{fig:rhohi} we compare our measurements of $\rho_{\rm HI}(z)$ to other high-redshift estimates: $z\sim 0 - 2$ DLA measurements \citep{Neeleman16,Shull17}, combined 21-cm \hi\ emission signal from star-forming galaxies at $z\approx 1$ \citep{Chowdhury20}, results from a survey of 21-cm \hi\ absorption systems at $z=0-2.74$ \citep{Grasha20}, and the recent compilation of $z\sim 2-5$ DLA measurements \citep[see][and references therein]{Peroux20,Walter20}. 
We correct the values by 20\% to account for the fraction of \hi\ contained in sub-DLAs and Lyman-limit systems \citep[e.g.,][]{Zafar13}, but do not include the contribution of helium as we are only interested in the amount of hydrogen in the form of \hi.

Additionally, we compare our results to evolutionary trends of $\rho_{\rm HI}(z)$, considering both the empirically determined $\rho_{\rm HI,DLA}(z) = (5.8\pm 0.3)\times 10^7 \times (1+z)^{0.57\pm 0.04}$ from the available DLA measurements \citep{Peroux20}, the global $\rho_{\rm HI}(z)$ predicted by hydrodynamical simulations using various prescriptions for the inherent UV background \citep{Maio22}, and the predictions for the evolution of the \hi\ gas {\em in} galaxies from the "sub-res models" of the L-Galaxies simulations \citep{Yates21}. The simulations by \citet{Maio22} mainly track the amount of the cosmic gas mass density in neutral, atomic form, showing significant decreases at $z\approx 6-8$ from the initial state due to the reionization phase transition. The L-Galaxies simulations described \citet{Yates21} on the other hand are by construction designed to only model the neutral gas inside galaxies. We further make a prediction for $\rho_{\rm HI,ISM}(z)$ in galaxies by combining the redshift evolution of the stellar mass density $\rho_\star (z)$ from \citet[][but here converted to a Chabrier IMF\footnote{Effectively we adopt $M_\star$(Chabrier) = $M_{\star} $(Salpeter)/1.5}]{Walter20} and the \hi\ gas mass excess $M_{\rm HI} / M_{\star} \propto (1+z)^{1.6\pm 0.2}$ derived in Sect.~\ref{sec:res}, as 
\begin{equation}\label{eq:rhohi}
    \rho_{\rm HI,ISM}(z) = \rho_\star (z) \times M_{\rm HI} / M_\star (z) ~.
\end{equation}
Propagating the uncertainties on the functions for $\rho_\star (z)$ and $M_{\rm HI} / M_\star (z)$, yields a relative uncertainty of $\approx 35\%$ on $\rho_{\rm HI,ISM}(z)$.
We find that the evolution of $\rho_{\rm HI}(z)$ predicted from simulations and the DLA measurements show significant excess over the inferred values of $\rho_{\rm HI}$ from the [\cii]-to-\hi\ relation at $z\gtrsim 5$ from this work. We observe, however, a striking match between our measurements and the evolutionary trend predicted for $\rho_{\rm HI,ISM}(z)$ from Eq.~\ref{eq:rhohi} and the L-Galaxies simulations by \citet{Yates21}. While both our measurements of $\rho_{\rm HI}$ and the function described by Eq.~\ref{eq:rhohi} rely on the [\cii]-to-\hi\ conversion factor, we emphasize that one approach adopts the independently-derived [\cii]-$158\mu$m LF, whereas the other is inferred from literature measurements of $\rho_\star (z)$ and the average redshift evolution of the \hi\ gas mass excess. This discrepancy indicates that the majority of the \hi\ gas probed by DLAs at $z\gtrsim 2$ is not associated with the star-forming regions of galaxies at this epoch. 

\begin{figure}[!t]
\centering
\includegraphics[width=9cm]{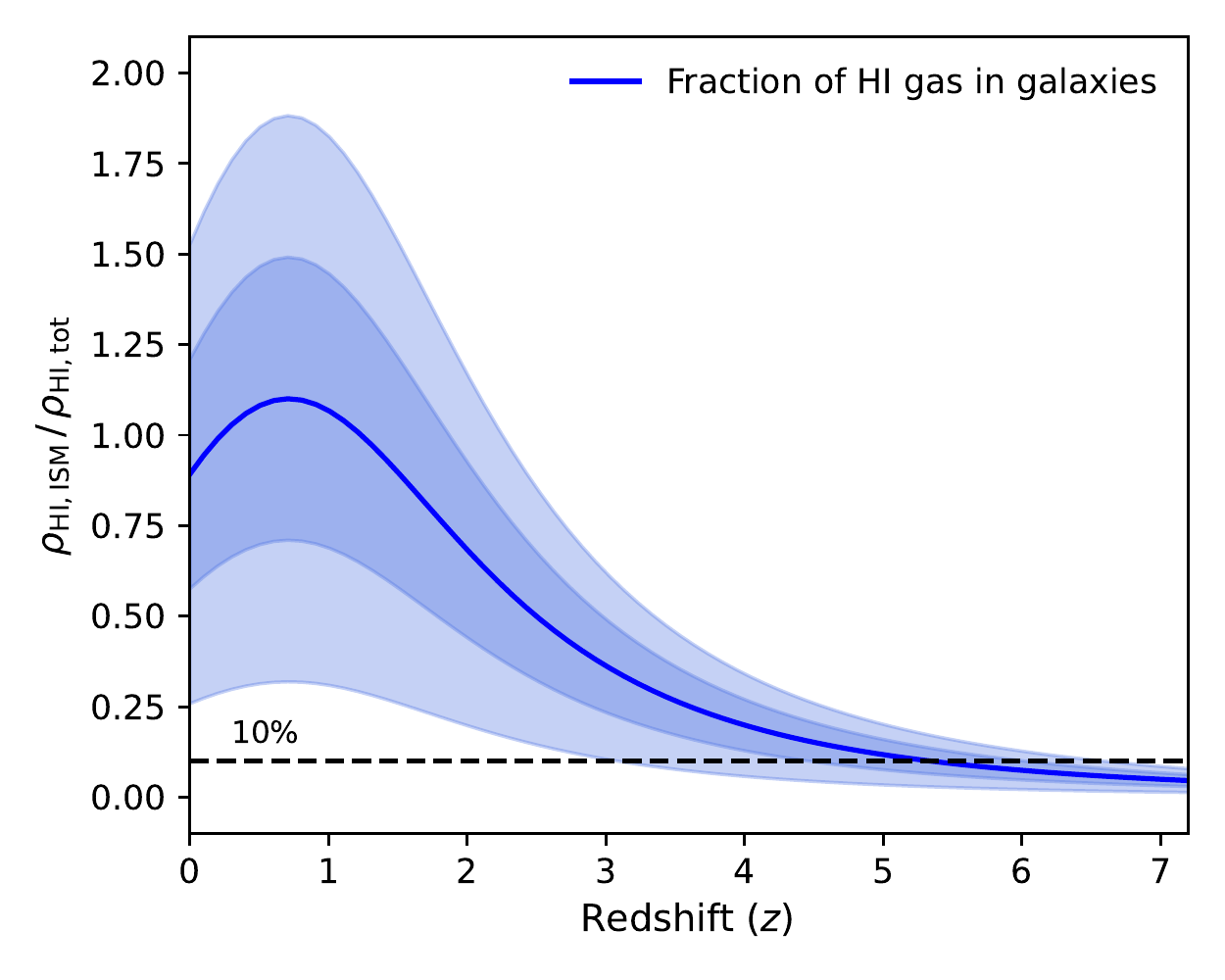}
\caption{Fraction of \hi\ in galaxies, $\rho_{\rm HI,ISM} / \rho_{\rm HI,tot}$, as a function of redshift. $\rho_{\rm HI,ISM}$ measures the \hi\ associated with star formation in the neutral ISM probed by [\cii], whereas $\rho_{\rm HI,tot}$ measures the total \hi\ both in the ISM and the extended halos of galaxies probed by random quasar sightlines. Solid blue lines shows the relative function and the dark- and light-shaded regions the $1\sigma$ and $2\sigma$ confidence intervals. At $z\lesssim 2$ most ($\gtrsim 70\%$) of the \hi\ gas has accreted into galaxies.}
\label{fig:rhocomp}
\end{figure}

We here propose a physical interpretation of this apparent discrepancy between $\rho_{\rm HI}$ measured from DLAs and via $\beta_{\rm [CII]}$. We argue that the former represents the {\em total} cosmic \hi\ gas mass density, as expected given the random quasar sightlines through galaxies and the correction for amount of \hi\ in sub-DLAs and Lyman-limit systems, where the latter reflects the amount of \hi\ contained {\em in} the neutral, star-forming ISM of galaxies. In this scenario, the evolutionary trend of $\rho_{\rm HI,ISM}(z)$ described by Eq.~\ref{eq:rhohi} represents the build-up and accumulation of \hi\ in galaxies, acting as the fundamental gas reservoir from which molecular gas and stars can be formed. On the other hand, the large excess of \hi\ gas measured by quasar absorption-line systems at $z\gtrsim 3$ hints that the bulk of \hi\ is located in the halos of galaxies at these redshifts, and thus not directly associated with star formation \citep[as also revealed by recent ALMA observations of quasar absorbers;][]{Neeleman19}.
This is also consistent with the recent simulations by \citet{Stern21}, in addition to \citet{Yates21} who found that at least $\approx 65\%$ of the \hi\ gas in the L-Galaxies, EAGLE, and TNG100 simulations have to be located in the outskirts of galaxies beyond $2\times$ the stellar half-mass radius at $z\gtrsim 3$.
We quantify this relative ratio and thereby the build-up of \hi\ in galaxies in Fig.~\ref{fig:rhocomp}. Here, $\rho_{\rm HI,tot}$ is described by the empirical relation $\rho_{\rm HI,DLA}(z)$ from DLA sightlines \citep{Peroux20} and $\rho_{\rm HI,ISM}$ is a measure of the \hi\ associated with star formation in the neutral ISM probed by [\cii] as described by Eq.~\ref{eq:rhohi}.
We find that at $z\approx 3$ and beyond, the majority of \hi\ ($\gtrsim 60\%$) is located in the halos or circumgalactic medium of galaxies. At $z\gtrsim 5$, only $\approx 10\%$ of \hi\ confined in the ISM of galaxies. This simple model further predicts that by $z\approx 2$, the bulk of the extended \hi\ gas reservoirs will have accreted onto the galaxy ISM, exhausting the available circum- or intergalatic gas. We would like to caution though that this is only a first preliminary study on large cosmic time- and spatial-scales, with potential biases related to sample statistics, or different methods to probe the \hi\ mass densities. We thus encourage complementary observations or simulations to further investigate and constrain this proposed scenario.

\section{Summary and outlook} \label{sec:conc}

We have here presented the first measurements of the neutral atomic hydrogen (\hi) gas content of star-forming galaxies during the epoch of reionization. This work is based on the sample of galaxies at $z= 6.5 - 7.8$ surveyed for [\cii]$-158\mu$m emission as part of the REBELS survey \citep{Bouwens22}. Combined with a recent calibration of the [\cii]-to-\hi\ conversion factor, $\beta_{\rm [CII]} = M_{\rm HI}/L_{\rm [CII]}$, we inferred the \hi\ gas masses of the individual galaxies in this survey based on the [\cii] luminosities, $L_{\rm [CII]}$. We found an average \hi\ gas mass excess of $M_{\rm HI}/M_\star \approx 10$, corresponding to an average \hi\ gas mass fraction of $f_{\rm HI} = M_{\rm HI}/(M_\star + M_{\rm HI}) \approx 90\%$. These results are consistent with an overall, gradual increase of $M_{\rm HI}/M_\star$ and $f_{\rm HI}$ with redshift. Further, they provide the first evidence that \hi\ is the dominant baryonic component by mass in galaxies from the epoch of reionization to $z\approx 2$.

We further used the $\beta_{\rm [CII]}$ conversion factor to make predictions for the cosmic \hi\ gas mass density in galaxies ($\rho_{\rm HI}$) at $z\approx 7$ based on recent estimates of the [\cii]-$158\mu$m luminosity function at this epoch. We found that $\rho_{\rm HI}$ increases by a factor of $\approx 10$ from $z\approx 7$ to $z\approx 2$, with $\rho_{\rm HI} = 7.1^{+6.4}_{-3.0}  \times 10^{6}\,M_\odot$\,Mpc$^{-3}$ at $z\approx 7$. We argued that this increase reflects the accretion and build-up of \hi\ onto galaxies, associated with the star-forming ISM as traced by [\cii]. Notably, this evolutionary trend shows a large deficit of \hi\ when compared to measurements from DLA sightlines and predictions of the global \hi\ gas content from hydrodynamical simulations at similar cosmic epochs. To explain this apparent discrepancy, we proposed a scenario where the DLA sightlines mainly probe \hi\ in the halos of galaxies at $z\gtrsim 2$, thereby not associated with the bulk of star formation. This indicates that the fraction and distribution of \hi\ in halos and in the ISM of galaxies are markedly different at low and high redshifts.

These results emphasizes the need to also take into account the neutral atomic gas-phase, which may be the dominant baryonic component of high-$z$ galaxies, instead of assuming that the molecular gas-phase is the sole contributor to $M_{\rm gas}$ as done previously \citep[e.g.,][]{Tacconi20,ForsterSchreiber20}. Neglecting \hi\ might affect dynamical studies, inferred gas depletion timescales and the overall assembly history of gas in galaxies and stellar-mass build-up. Recently, \citet{Walter20} demonstrated that the growth in stellar mass cannot be accounted for by the decrease in the cosmic H$_2$ gas in galaxies, requiring significant infall of additional pristine gas. Here we constrained this exact process, disentangling the amount of \hi\ in and outside of galaxies through cosmic time. The \hi\ gas in galaxies associated with the star-forming ISM acts as the transition state between the circumgalatic hot halo gas and the cool, molecular gas-phase providing the fuel for star formation.

With the advent of the next generation radio facilities such as the Square Kilometre Array (SKA) it will soon be possible to survey \hi\ directly through the 21-cm transition from individual galaxies beyond what is currently achievable \citep[][]{Blyth15}. SKA pathfinder telescopes such as MeerKAT have already shown great promise in extending the redshift range for which the 21-cm \hi\ emission can be detected in individual galaxies \citep{Maddox21}. However, even with the SKA it will still be challenging to detect \hi\ directly from galaxies much beyond $z>0.5$. Statistical approaches averaging the spectra of known galaxies \citep{Chowdhury20} or intensity mapping \citep{Santos15} might potentially alleviate this technical limitation, but still only provide an average census of \hi\ in galaxies at the surveyed redshifts. Using [\cii] as a proxy for \hi\ is therefore the current best alternative to probe the neutral atomic gas-phase in galaxies beyond the local Universe and out to the epoch of reionization.

\section*{Acknowledgements}

We would like to thank the referee for a detailed and constructive review, that has greatly improved the presentation of this paper. 
K.E.H. would like to thank Darach Watson, Johan Fynbo, and Peter Laursen for insightful discussions during the early stages of this work. K.E.H. would further like to thank Marcel Neeleman and Fabian Walter for their hospitality at MPIA and the fruitful discussions on the implications of this work.
This paper makes use of the following ALMA data: ADS/JAO.ALMA\#2019.1.01634.L. ALMA is a partnership of ESO (representing its member states), NSF (USA) and NINS (Japan), together with NRC (Canada), MOST and ASIAA (Taiwan), and KASI (Republic of Korea), in cooperation with the Republic of Chile. The Joint ALMA Observatory is operated by ESO, AUI/NRAO and NAOJ.
K.E.H. acknowledges support from the Carlsberg Foundation Reintegration Fellowship Grant CF21-0103.
The Cosmic Dawn Center (DAWN) is funded by the Danish National Research Foundation under grant No. 140.
M.A. acknowledges support from FONDECYT grant 1211951 "ANID+PCI+INSTITUTO MAX PLANCK DE ASTRONOMIA MPG 190030", "ANID+PCI+REDES 190194" and ANID BASAL project FB210003.
A.F. and A.P. acknowledge support from the ERC Advanced Grant INTERSTELLAR H2020/740120. Generous support from the Carl Friedrich von Siemens-Forschungspreis der Alexander von Humboldt-Stiftung Research Award is kindly acknowledged (A.F.). 
Y.F. acknowledges support from NAOJ ALMA Scientific Research Grant number 2020-16B.
P.D. acknowledges support from the European Research Council's starting grant ERC StG-717001 (``DELPHI"), from the NWO grant 016.VIDI.189.162 (``ODIN") and the European Commission's and University of Groningen's CO-FUND Rosalind Franklin program.
H.I. acknowledges support from NAOJ ALMA Scientific Research Grant Code 2021-19A. H.I. acknowledges support from JSPS KAKENHI Grant Number JP19K23462.
R.S. acknowledges support from a STFC Ernest Rutherford Fellowship (ST/S004831/1).

\section*{Data availability statement} 

Source codes for the figures and tables presented in this manuscript are available from the corresponding author upon reasonable request.\\


\bibliography{ref}
\bibliographystyle{aasjournal}

\end{document}